\documentclass[3p, times, twocolumn, sort&compress]{elsarticle}
\usepackage{ecrc}
\usepackage{color}
\usepackage{amssymb}
\usepackage[english]{babel}
\usepackage{amsfonts}
\usepackage{amsmath}
\usepackage{array}
\usepackage{float}
\usepackage[version=3]{mhchem}
\usepackage{textcomp}
\usepackage{placeins}
\usepackage{footnote}
\usepackage[para]{threeparttablex}
\makesavenoteenv{tabular}
\makesavenoteenv{table}
\newcolumntype{x}[1]{>{\centering\arraybackslash\hspace{0pt}}p{#1}}

\volume{00}
\firstpage{1}
\journalname{Chemical Physics}
\runauth{J. Schulze et al.}
\jid{cp}
\jnltitlelogo{Chemical Physics}
\CopyrightLine{2011}{Published by Elsevier Ltd.}
\begin{document}
\def\cm{cm$^{-1}$}
\newcommand*{\eref}[1]{Eq.~(\plainref{#1})}
\newcommand*{\fref}[1]{Fig.~\plainref{#1}}
\newcommand*{\tref}[1]{Tab.~\plainref{#1}}
\newcommand{\onlinecite}[1]{\hspace{-1 ex} \nocite{#1}\citenum{#1}} 
\newcommand*{\bref}[1]{Ref.~\onlinecite{#1}}
\newcommand*{\bra}[1]{\langle #1 |}
\newcommand*{\ket}[1]{| #1 \rangle}

\begin{frontmatter}
\title{The Effect of Site-Specific Spectral Densities on the High-Dimensional Exciton-Vibrational Dynamics in the FMO Complex}
%-------------------------
%
\author{Jan Schulze}
\address
{Institut f\"ur Physik, Universit\"at Rostock, Albert-Einstein-Str. 23-24, D-18059 Rostock, Germany}
\author{Mohamed F. Shibl\fnref{myfootnote}}\fntext[myfootnote]{Permanent address: Faculty of Science, Department of Chemistry, Cairo University, Giza, Egypt}
\address{Gas Processing Center, College of Engineering, Qatar University, P.O. Box 2713, Doha, Qatar}
\author{Mohammed J. Al-Marri}
\address{Gas Processing Center, College of Engineering, Qatar University, P.O. Box 2713, Doha, Qatar}
\author{Oliver K\"uhn}
\ead{oliver.kuehn@uni-rostock.de}
\cortext[cor1]{Corresponding author}
\address
{Institut f\"ur Physik, Universit\"at Rostock, Albert-Einstein-Str. 23-24, D-18059 Rostock, Germany}
\begin{abstract}
	The coupled exciton-vibrational dynamics of a three-site model of the FMO complex is investigated using the Multi-layer Multi-configuration Time-dependent Hartree (ML-MCTDH) approach. Emphasis is put on the effect of the spectral density on the exciton state populations as well as on the vibrational and vibronic non-equilibrium excitations. Models which use either a single or site-specific  spectral densities are contrasted to a spectral density adapted from experiment. For the transfer efficiency, the total integrated Huang-Rhys factor is found to  be more important  than details of the spectral distributions. However, the latter are relevant for the obtained non-equilibrium vibrational and vibronic distributions and thus influence the actual pattern of population relaxation.
\end{abstract}

\begin{keyword} 
Frenkel excitons \sep exciton-vibrational coupling \sep   quantum dynamics \sep  photosynthesis \sep FMO complex
\end{keyword}
\end{frontmatter}
\cleardoublepage
\newpage

\section{Introduction}
%
%Disorder has a vital role for the realization of highly efficient and directed excitation energy transfer in photosynthesis. Commonly, it is distinguished into dynamic and static disorder according to the relation of the time scales of fluctuations of environmental degrees of freedom (DOFs) and the relevant system's dynamics. Spectral signatures are homogeneous and inhomogeneous broadening as expressed in Kubo's famous lineshape model. In addition photosynthetic antennae exhibit a heterogeneous energy level structure of their pigment pools, realized by virtue of different electrostatic environments. It is essential for downward energy transfer as exemplified, e.g., by the Fenna-Matthews-Olson (FMO) complex of cyanobacteria.

The spectral density (SD) is central to the theory of  dissipative quantum dynamics~\cite{may11}. It describes the coupling of the relevant system to particular modes of the environmental bath. There is a number of model SDs (Ohmic, Debye-Drude or Multi-Mode Brownian Oscillator)~\cite{grabert88_115,weiss93,mukamel95}, whose general influence on the dynamics of model systems has been extensively studied. The actual definition of the SD is linked to an assumption concerning the system-bath coupling. For vibrational dynamics, the Caldeira-Leggett model, i.e. a bilinear form in system and bath coordinates, is typically assumed, although its applicability in general has recently been challenged~\cite{gottwald15_2722}. For problems involving an electronic excitation coupled to nuclear dynamics, the Huang-Rhys (HR) model is commonly applied. It assumes that vibrational degrees of freedom (DOFs) are described in harmonic approximation with the equilibrium positions of the oscillators being linearly shifted upon electronic excitation~\cite{mukamel95}. Recently,  SDs beyond simple models have attracted considerable attention in the context of Frenkel exciton dynamics in photosynthetic light-harvesting complexes. Here, the electronic excitation of the chlorophyll molecules is coupled to both, intramolecular and protein vibrations. While the SD for the latter is essentially structureless and often well described by model functions, intramolecular vibrations give rise to distinct features in the SD, whose spectral positions and weights might be relevant for the exciton dynamics~\cite{christensson12_7449}. 

Under the assumptions of the HR model, SDs can in principle be reconstructed from spectroscopic data such as site-selective fluorescence~\cite{jankowiak11_4546}.  For the widely discussed Fenna-Matthews-Olson (FMO) complex of cyanobacteria, Wendling et al.~\cite{wendling00_5825} have determined a SD by focussing on the lowest energetic bacteriochlorophyll $a$ (BChl~$a$) pigment at 4~K. Although their assumption that this particular BChl~$a$ molecule is electronically decoupled from the other BChl~$a$ molecules of the complex has been critically discussed~\cite{schulze14_045010}, the Wendling SD has become a standard for the discussion of FMO dynamics~\cite{adolphs06_2778,kreisbeck12_2828,schulze15_6211,schulze16_185101}. In Ref.~\cite{adolphs06_2778} the low-frequency phonon part had been found to be rather similar to that of the B877 monomer complex studied in Ref.~\cite{renger02_9997}. However, the Wendling SD, in contrast to the bare  phonon wing, contains structured features due to discrete vibrations. In Ref.~\cite{adolphs06_2778} this effect was modeled by adding an isolated delta-like peak to the SD. Such sharp features are a notorious problem for density matrix approaches to the dynamics. It can be circumvented by including the related vibrational mode into the relevant system~\cite{kuhn96_99,renger96_15654}. For the case of the FMO complex, this approach has been used to perform path integral~\cite{nalbach15_022706} and Quantum Master equation~\cite{liu16_272} simulations.

The computational determination of SDs for specific pigment-protein complexes usually employs sampling of the fluctuations of local electronic energy gaps using ground state equilibrium classical molecular dynamics. In a pioneering work, Schulten and coworkers have calculated the SD for BChl~$a$ in the light-harvesting antenna LH2 of purple bacteria~\cite{damjanovic02_031919}. Due to the limited trajectory length only the high-frequency part of the SD was accessible. Concerning the FMO complex there are essentially as many different SDs as there are published papers on this topic, although most of them agree in gross features. For instance, Kleinekath\"ofer and coworkers have determined site-specific FMO SDs using the semiempirical ZINDO/S approach to calculate electronic excitation energies~\cite{olbrich11_1771,olbrich11_8609}. A comparison of the effect of different force fields and electronic structure methods has been provided in Ref.~\cite{wang15_25629}. Further, the use of the classical approximation has been scrutinized in Ref.~\cite{valleau12_224103}. A different strategy has been followed by Renger et al., who used the shifted harmonic oscillator model directly by employing a normal mode analysis of the pigment-protein complex~\cite{renger12_14565}. The latest SD comes from the group of Coker et al.\ \cite{rivera13_5510,lee16_3171} and will also be used in the present work. The Coker SD combines both ideas mentioned above, i.e.\ the phonon wing is modeled using general gap correlation functions, whereas for the intramolecular vibrations a harmonic approximation is assumed~\cite{rivera13_5510,lee16_3171}.

In view of the many different FMO SDs, the question arises whether the details really matter for the dynamics of excitation energy transfer. In other words, are there any vibrational mode specific effects in a system as complicated as the FMO complex? Previously, we have shown that, in principle, an answer can be provided based on the propagation of the full exciton-vibrational wavepacket \cite{schulze15_6211,schulze16_185101}, which becomes possible by using the ML-MCTDH approach~\cite{meyer90_73,beck00_1,meyer03_251,meyer11_351,wang03_1289,manthe08_164116,vendrell11_044135}. Given an exciton Hamiltonian and a discretized SD, ML-MCTDH provides a numerical solution to the time-dependent Schr\"odinger equation, whose convergence to a desired accuracy can be monitored.  

In the present contribution, ML-MCTDH is applied to FMO dynamics using different SDs, i.e. the Wendling~\cite{wendling00_5825} and the Coker SD~\cite{lee16_3171}. This will allow us to highlight the sensitivity of the dynamics with respect to the details of the SD model. The paper starts with a brief outline of Frenkel exciton theory and ML-MCTDH in Section \ref{sec:methods}. Here, we will also introduce the different SD models. Results of numerical simulations are discussed in Section \ref{sec:results} and a summary is provided in Section \ref{sec:summary}.
\section{Theoretical Methods}
\label{sec:methods}
\subsection{Exciton-Vibrational Hamiltonian} %------------------------------------------------
The Frenkel exciton Hamiltonian describes an aggregate with $ N_{\rm agg} $ sites (site index $m$), each site having the excitation energy $E_m$, and different sites being coupled by the Coulomb interaction $J_{mn}$\cite{may11}
\begin{equation} 
H_{\rm ex} =\sum_{m,n=1}^{N_{\rm agg}}(\delta_{mn} E_m + J_{mn})\ket{m}\bra{n} \, .
\end{equation}
Here, we used the Frenkel one-exciton states $\ket{m}=\ket{e_m}\prod_{n\ne m}\ket{g_n}$, which are defined in terms of the local 
electronic ground, $\ket{g_m}$, and excited, $\ket{e_m}$, states. For the site energies and Coulomb interactions, we will use the FMO values reported by Moix et al.~\cite{moix11_3045}. They are based on a  combination of  site energies obtained from quantum chemical/electrostatic calculations~\cite{schmidtambusch11_93} and Coulomb couplings described within the dipole-dipole approximation. Previously, it has been shown that the dynamics is essentially confined to the sites 1 to 3~\cite{schulze16_185101,moix11_3045}. This justifies the restriction to these three sites in the following. Thus the Hamiltonian matrix   is given by  (in units of \cm, off-set is 12195 \cm)~\cite{moix11_3045}:
\begin{eqnarray}
\label{eq:hmat}
\mathbf{H}_{\mathrm{ex}}&=&\left(  
    \begin{array}{ccc}
      310&{-98}&6\\
      {-98}&230&{30}\\
      6&{30}&0\\
    \end{array}                    
  \right)\,.
\end{eqnarray}
Note that the labeling of the sites follows the structure of the Hamiltonian matrix, e.g., site $m=3$ is the energetically lowest site, which is connected to the cytoplasmic membrane containing the  reaction center complex.  

\begin{figure}[t]
\centering
\includegraphics[width=0.95\columnwidth]{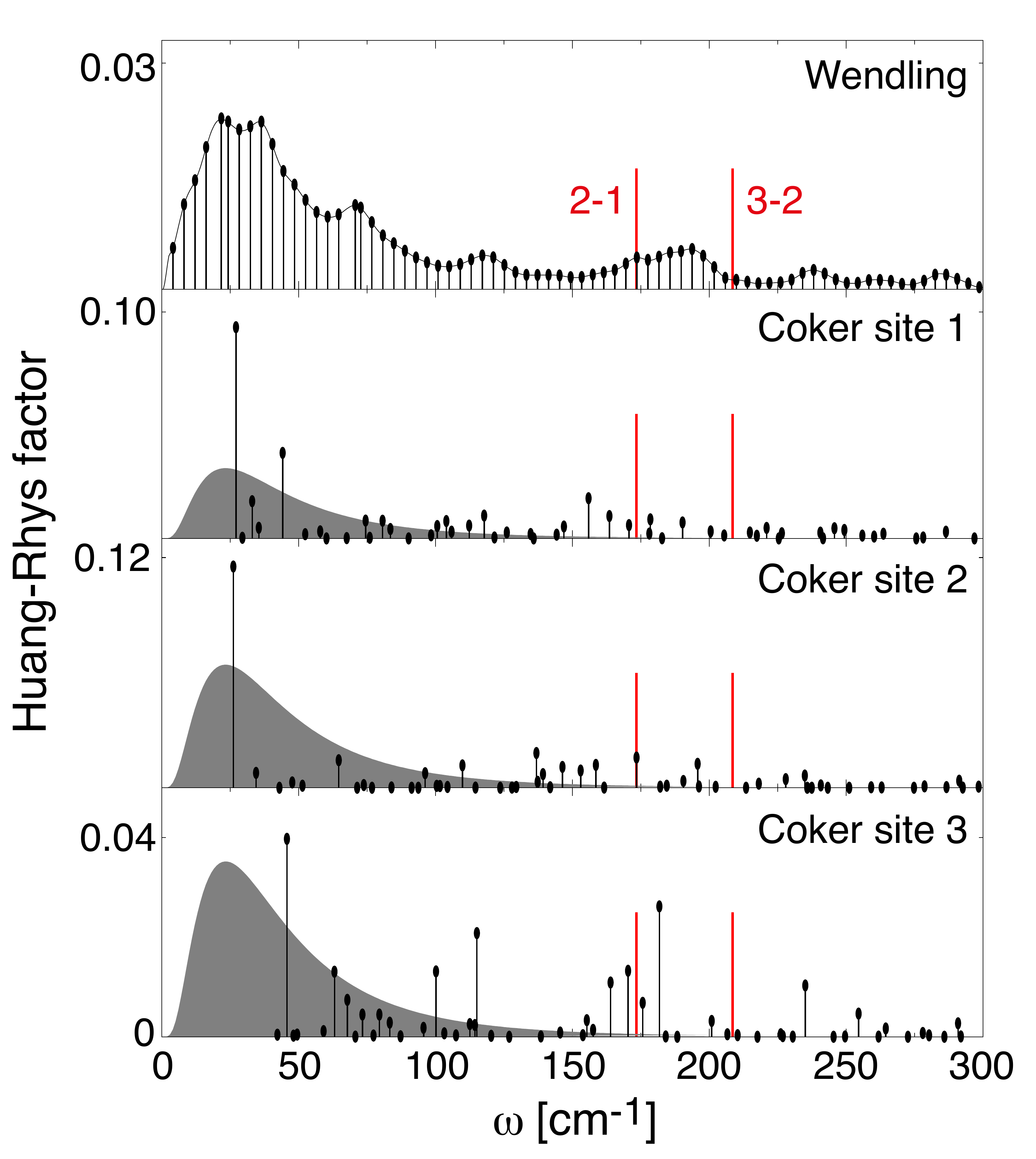}
\caption{
Distribution of Huang-Rhys factors in the SD according to the experiment of Wendling et al. \cite{wendling00_5825} and the site-specific calculations of Coker et al. \cite{lee16_3171}. In the lower three panels the SDs are separated into a phonon wing (grey) and intramolecular mode contributions (sticks). The transition energies between adiabatic states are given as red bars $(E_{\alpha=3}-E_{\alpha=2}=208$~\cm{} and $E_{\alpha=2}-E_{\alpha=1}=173$~\cm).
}
\label{fig:sd}     
\end{figure}

Diagonalization of this matrix yields the (in the following called adiabatic) one-exciton eigenstates $|\alpha \rangle = \sum_m c_{m}(\alpha) |m\rangle$ with energies $E_\alpha$. The related transition energies are given in Fig.~\ref{fig:sd}. The decompositions
 into the local (in the following called diabatic) states $\ket{m}$ are as follows (in  order of decreasing energy): ${\bf c}(3)=(-0.83,  0.56,  0.03)$, ${\bf c}(2)=(0.56, 0.81, 0.16)$, and ${\bf c}(1)=(-0.06, -0.15, 0.99)$.
 
%
%ffffffffffffffffffffffffffffffffffffffffffffffffffffffffffffffffffffffff
%%ffffffffffffffffffffffffffffffffffffffffffffffffffffffffffffffffffffffff
%
The local vibrations at site $m$ are described in harmonic approximation by the set of dimensionless normal mode coordinates $ \{ Q_{m,\xi} \} $ with frequencies $ \{ \omega_{m,\xi} \} $, i.e. the vibrational Hamiltonian reads
\begin{equation}
H_{\rm vib} =\sum_m \sum_{\xi \in m} h_{m,\xi}\, ,
\end{equation}
with the harmonic oscillator Hamiltonian
\begin{equation}
	h_{m,\xi}=\frac{\hbar\omega_{m,\xi}}{2} \left( - \frac{\partial^2}{\partial Q_{m,\xi}^2}+ Q_{m,\xi}^2\right) \,.
\end{equation}

EVC is accounted for within the linearly shifted oscillator  model, i.e.
\begin{equation}
H_{\rm ex-vib} = \sum_m \sum_{\xi \in m} \hbar \omega_{m,\xi} \sqrt{2 S_{m,\xi}}Q_{m,\xi} \ket{m}\bra{m}\, .
\end{equation}
%U_{m,a} & = & E_{m,a}    +
The coupling of a particular mode to the electronic transition is characterized by the Huang-Rhys (HR) factor $S_{m,\xi}$.

Frequencies and HR factors can be obtained from the SD, $J_m(\omega)$, of the monomeric BChl~$a$ molecule~\cite{may11}
\begin{equation}
\label{eq:jomega}
	J_m(\omega) = A \sum_{\xi\in m}   S_{m,\xi} \delta(\omega-\omega_{m,\xi})\, ,
\end{equation}
where $A$ is a constant that will be used to adjust the total HR factor for site $m$ for a finite discretization according to $S_{\rm tot}=A^{-1}\int d\omega J_{m}(\omega) = \sum_{\xi \in m} S_{m,\xi}$.

 Since the reported SDs differ considerably, we have used the experimentally determined SD of Wendling et al.~\cite{wendling00_5825} in our previous investigation (cf. Fig.~\ref{fig:sd})~\cite{schulze15_6211,schulze16_185101}.  In the present study, the Wendling SD will be taken as a reference and will be called \textit{model I}. Specifically, it is discretized into 74 modes within the interval $[2:300]$ \cm{} as shown in Fig.~\ref{fig:sd}.  The amplitudes of the individual HR factors have been adjusted homogeneously via the constant $A$ such as to  preserve the total HR factor, $S_{\rm tot}=0.42$, upon summation. 
 
 The results of model I will be compared to those obtained using the site-specific Coker SDs of Ref.~\cite{lee16_3171}, called \textit{model II}. In Fig.~\ref{fig:sd} these SDs are decomposed into a phonon wing and a discrete intramolecular part. The former has been fitted to a log-normal distribution, i.e. ($S_{\rm ph}=(0.33, 0.68, 0.37)$ for sites $(1,2,3)$ and $\sigma=0.7$, $\omega_{\rm c}=38$~\cm)
 \begin{equation}
 	J_{\rm ph}(\omega)=\frac{\pi S_{\rm ph}\omega}{\sqrt{2\pi}\sigma}\exp\left\{-\frac{[\ln(\omega/\omega_c)]^2}{2\sigma^2}\right\} \, .
 \end{equation}
 Note that in Ref.~\cite{rivera13_5510} a different definition of the SD had been used. The present $J_{\rm ph}$ are chosen such as to give the same reorganisation energies. The $J_{\rm ph}$ have been discretized in the interval $[2:160]$ \cm{} into 32 modes. The intramolecular part was taken directly from Ref.~\cite{lee16_3171}. This results in a total of 81 modes for each site. In the Coker model II the total HR factors are site-specific, i.e. $S_{\rm tot}=(0.64, 0.96, 0.58)$ for sites $(1, 2, 3)$. Notice that these values exceed the one extracted from the experimental data (0.42) by Wendling et al.~\cite{wendling00_5825}. Therefore, we will consider \textit{model III}, where the mode structure of model II is kept, but all HR factors are uniformly scaled to the experimental value $S_{\rm tot}=0.42$. Finally, \textit{model IV} is designed such that all sites share the scaled Coker SD of site 3, i.e. the mode structure is uniform but different from model I.

As a final note in caution, we would like to point out that the discretization leads to a recurrence time, $T_{\rm rec}=2\pi/\Delta \omega$,  of about 8~ps for the monomer in case of model I. For the Coker SD, the situation is more complicated due to the dominance of a few discrete peaks in the intramolecular part. Indeed, this is the reason why we will restrict the propagation time to 1~ps. Beyond this time, effects of recurrences in the population dynamics start to appear (not shown), which can be considered as an artifact of the model.
\subsection{Quantum Dynamics}%%%%%%%%%%%%%%%%%%%%%%%%%%%%%%%%%%%%%%%%%%%%%%%%%%%%%%%
The time-dependent Schr\"odinger equation will be solved employing the ML-MCTDH method (for a review, see Ref.~\cite{meyer11_351}). The state vector is expanded into the local exciton basis according to
\begin{eqnarray}
	|\Psi({\bf Q};t) \rangle=\sum_{m} \chi_{m }({\bf Q};t) \, |m \rangle \,.
\end{eqnarray}
The  nuclear coordinates are comprised into the $D=N_{\rm agg}\times N_{\rm vib}$ dimensional vector $\mathbf{ Q}$.  Here, $N_{\rm vib}$ is the number of modes per site, which is assumed to be site-independent. The nuclear wave function is  expanded into MCTDH form 
\begin{equation}
\label{eq:psiMCTDH}
\chi_m(\mathbf{ Q},t) = \sum_{j_1 \ldots j_D}^{{n_{j_1} \ldots n_{j_D}}}
C^{(m)}_{j_1,\ldots,j_D}(t) \phi^{(m)}_{j_1}(Q_1;t) \ldots \phi^{(m)}_{j_D}(Q_{D};t) \, .
\end{equation}
Here, the $C^{(m)}_{j_1,\ldots,j_D}(t)$ are the time-dependent expansion coefficients weighting the contributions of the different Hartree products, which are composed of $n_{j_{k}}$ single particle functions (SPFs), $\phi^{(\alpha)}_{j_k}(Q_k;t)$, for the $k$th degree of freedom in state $|m\rangle$.  In ML-MCTDH the SPFs themselves describe multi-dimensional logical coordinates that are expanded into MCTDH form~\cite{wang03_1289,manthe08_164116,vendrell11_044135}. This yields a nested set of expansions that can be represented by so-called ML-MCTDH trees~\cite{manthe08_164116}.  The particular choice of this tree  strongly influences the required numerical effort~\cite{vendrell11_044135,schroter15_1}; for applications to coupled electron-vibrational dynamics, see also Refs.~\cite{meng12_134302,meng13_014313}. In the following simulations we use a grouping according to the magnitude of the HR factor and frequency as detailed in Ref.~\cite{schulze15_6211}. 

Wave packet propagations have been performed using the Heidelberg program package \cite{mctdh85}. The initial conditions has been a vertical Franck-Condon transition at site $m=1$  (with respect to a Hartree product ground state composed of non-shifted harmonic oscillators) and the propagation time was 1~ps. Convergence of the ML-MCTDH setup has been monitored by means of the grid size, the precision of the integrator, and the  natural orbital populations \cite{beck00_1}. The largest  population of the least occupied natural orbital was typically $\sim 10^{-3}$.

The quantum dynamics will be characterized by means of the  exciton  populations either in site (diabatic) 
$P_m(t) = \langle \Psi(t)| m \rangle \langle m | \Psi(t) \rangle $ or in eigenstate (adiabatic)  $P_\alpha(t) = \langle \Psi(t)| \alpha \rangle \langle \alpha | \Psi(t) \rangle $ representation. The latter are obtained from the propagated state vector via $P_\alpha(t)=\sum_{mn} c_m(\alpha)c^*_n(\alpha) \langle \Psi(t)| m\rangle\langle n| \Psi(t) \rangle$.

Vibrational excitation in the electronic ground and excited state will be called vibrational and vibronic excitation, respectively. The energy of the vibrational excitation at site $m$ follows from the expectation value of the  operator
\begin{equation}
\label{eq:vibra}
H^{\rm (vibra)}_m=\sum_{\xi \in m}h_{m,\xi}(1-|m\rangle\langle m|) \,,
\end{equation}
which gives the vibrational energy irrespective which site of the aggregate is electronically excited.

%
%ffffffffffffffffffffffffffffffffffffffffffffffffffffffffffffffffffffffff
\begin{figure*}[t]
\centering
\includegraphics[width=0.775\textwidth,angle=0]{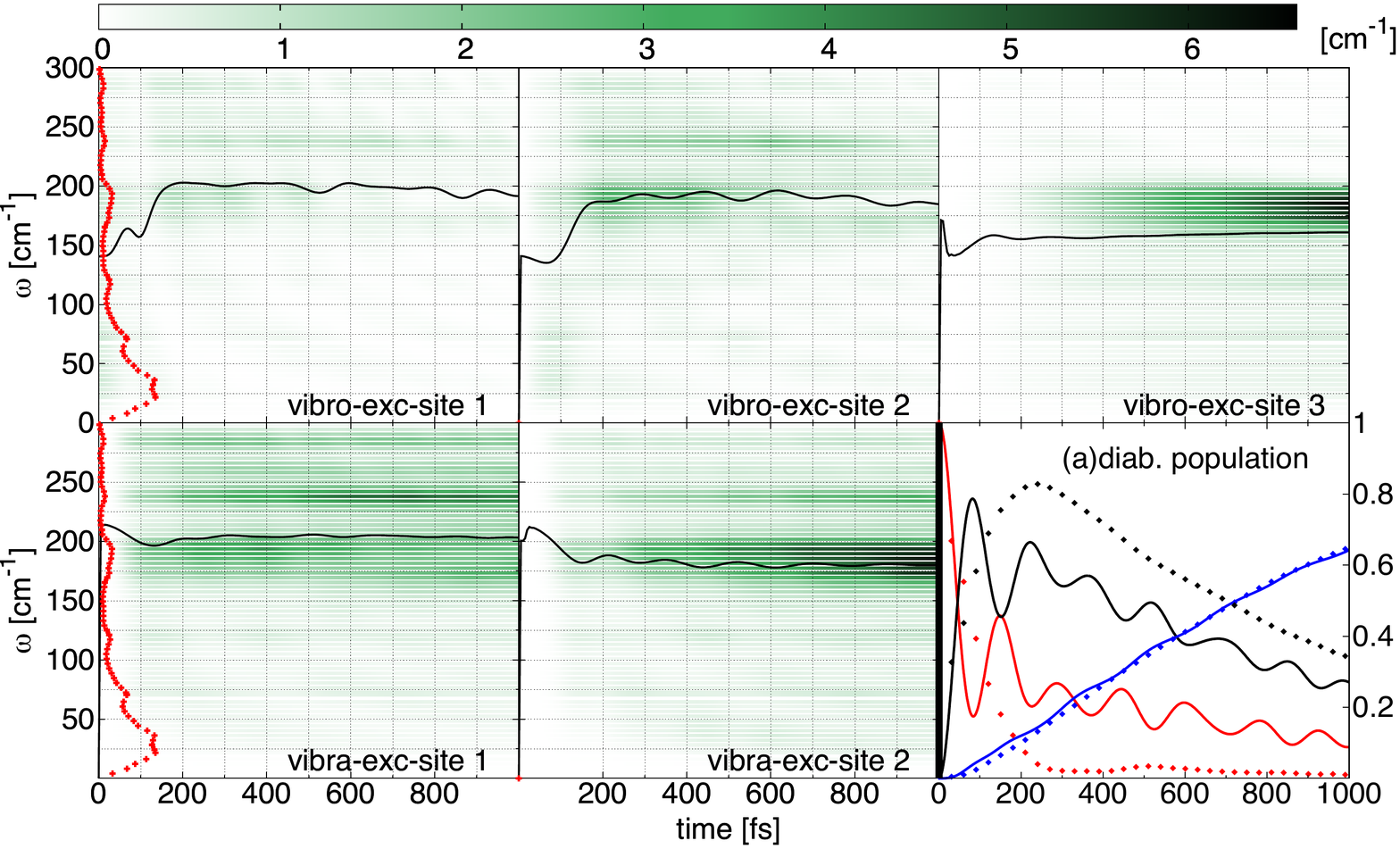}
\caption{(color online) Energy expectation values of the three-site model with the Wendling SD (shown next to the right axes) for the vibrational (lower panel) and vibronic (upper panel) Hamiltonian according  to Eqs.~(\ref{eq:vibra}) and (\ref{eq:vibro}), respectively (model I). The non-equilibrium vibrational excitation of site 3 is negligible (not shown). The black lines represent the averaged excitation energy according to Eq.~(\ref{eq:em}). The lower right panel shows the population dynamics in the diabatic ($P_m(t)$, solid) and adiabatic ($P_\alpha(t)$, dotted) basis (states 1, 2, and 3 correspond to the red, black and blue line, respectively, in the diabatic basis and vice versa in the adiabatic one.)
}
\label{fig:wendling}     
\end{figure*}
%%ffffffffffffffffffffffffffffffffffffffffffffffffffffffffffffffffffffffff

%%ffffffffffffffffffffffffffffffffffffffffffffffffffffffffffffffffffffffff
\begin{figure*}[b]
\centering
\includegraphics[width=0.775\textwidth,angle=0]{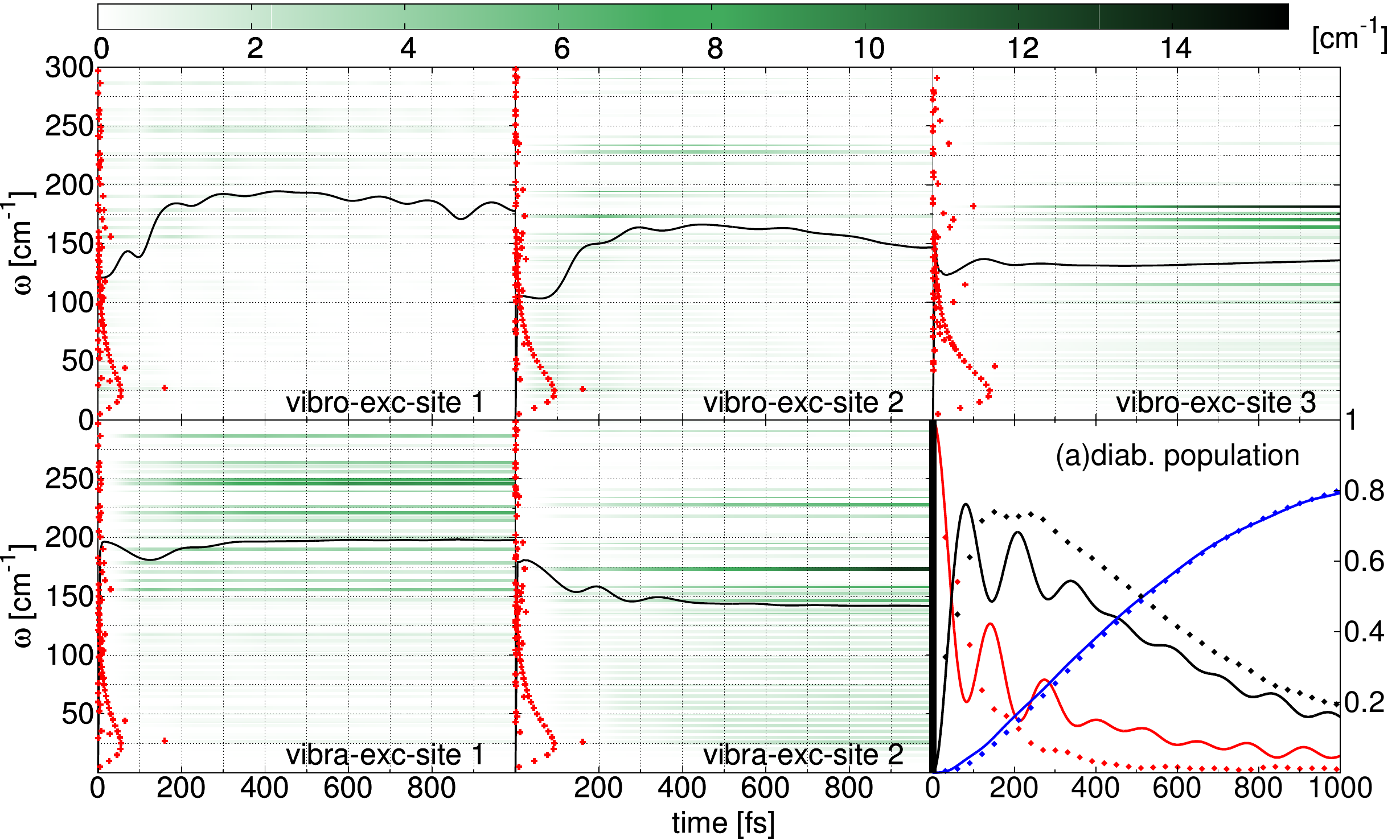}
\caption{
(color online) Same as Fig.~\ref{fig:wendling}, but for the full Coker model (model II).
}
\label{fig:coker_full}     
\end{figure*}
%%ffffffffffffffffffffffffffffffffffffffffffffffffffffffffffffffffffffffff
\cleardoublepage
%%ffffffffffffffffffffffffffffffffffffffffffffffffffffffffffffffffffffffff
\begin{figure*}[t]
\centering
\includegraphics[width=0.55\textwidth,angle=90]{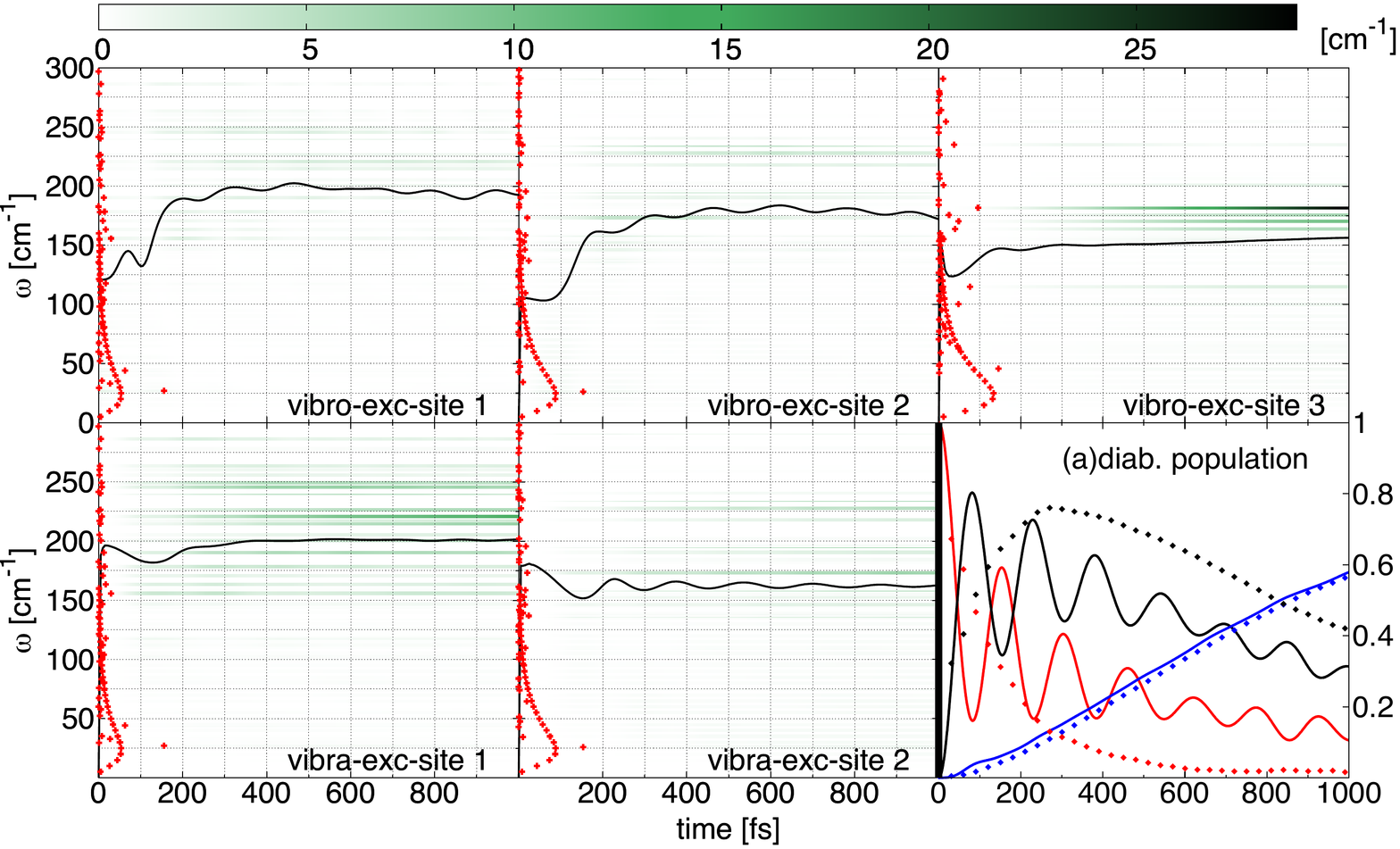}
\caption{
(color online) Same as Fig.~\ref{fig:wendling}, but for the rescaled Coker model (model III).
}
\label{fig:coker_rescaled}     
\end{figure*}
%ffffffffffffffffffffffffffffffffffffffffffffffffffffffffffffffffffffffff

%%ffffffffffffffffffffffffffffffffffffffffffffffffffffffffffffffffffffffff
\begin{figure*}[b]
\centering
\includegraphics[width=0.55\textwidth,angle=90]{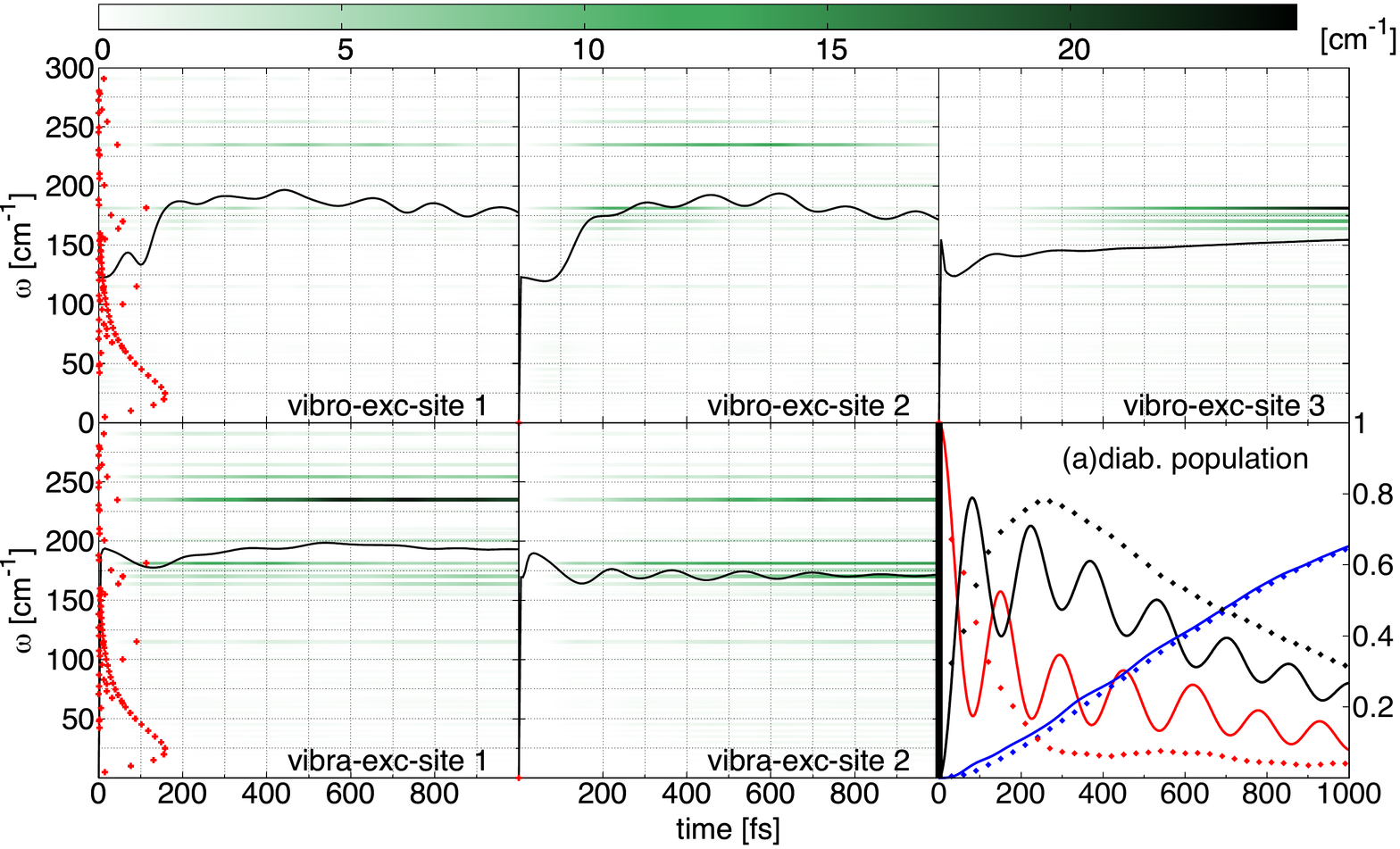}
\caption{
(color online) Same as Fig.~\ref{fig:wendling}, but for the rescaled Coker model using only the site 3 SD (model IV).
}
\label{fig:coker_rescaled_site3}     
\end{figure*}
%ffffffffffffffffffffffffffffffffffffffffffffffffffffffffffffffffffffffff
\cleardoublepage
The vibronic energy at site $m$ is defined by the expectation value of the operator
\begin{eqnarray}
\label{eq:vibro}
H^{\rm (vibro)}_m &=&
\sum_{\xi \in m} \left( h_{m,\xi} +\omega_{m,\xi} \sqrt{2 S_{m,\xi}}  Q_{m,\xi} \right) |m\rangle\langle m|\,. \nonumber\\&& 
\end{eqnarray}

As a global measure of the vibrational and vibronic excitation we will calculate

\begin{equation}
\label{eq:em}
{\mathcal E}_m (t)=\hbar \sum\limits_{\xi \in m}^{N_{\rm vib}} p(t,\omega_{m,\xi})\omega_{m,\xi}	
\end{equation}
where $ p(t,\omega_{m,\xi})$ is the distribution of excitation energies for the modes in a particular electronic state at a given time $t$. It is calculated from the expectation values of the terms contributing to the sums in Eqs.~(\ref{eq:vibra}) and (\ref{eq:vibro}). Since this distribution changes with the state populations, it is normalized to unity at each time step.

Further, we will inspect the local energy gaps defined as
\begin{eqnarray}
\label{eq:gap}
	\Delta_m(t)&=& E_m + \sum_{\xi \in m} \hbar \omega_{m,\xi} \sqrt{2 S_{m,\xi}}\nonumber\\
	&\times & \int d{\bf Q}\,  \chi_{m }^*({\bf Q};t)\,  Q_{m,\xi} \, \chi_{m }({\bf Q};t) \,.
\end{eqnarray}

%ffffffffffffffffffffffffffffffffffffffffffffffffffffffffffffffffffffffff
\begin{figure}[t]
\includegraphics[width=0.475\textwidth]{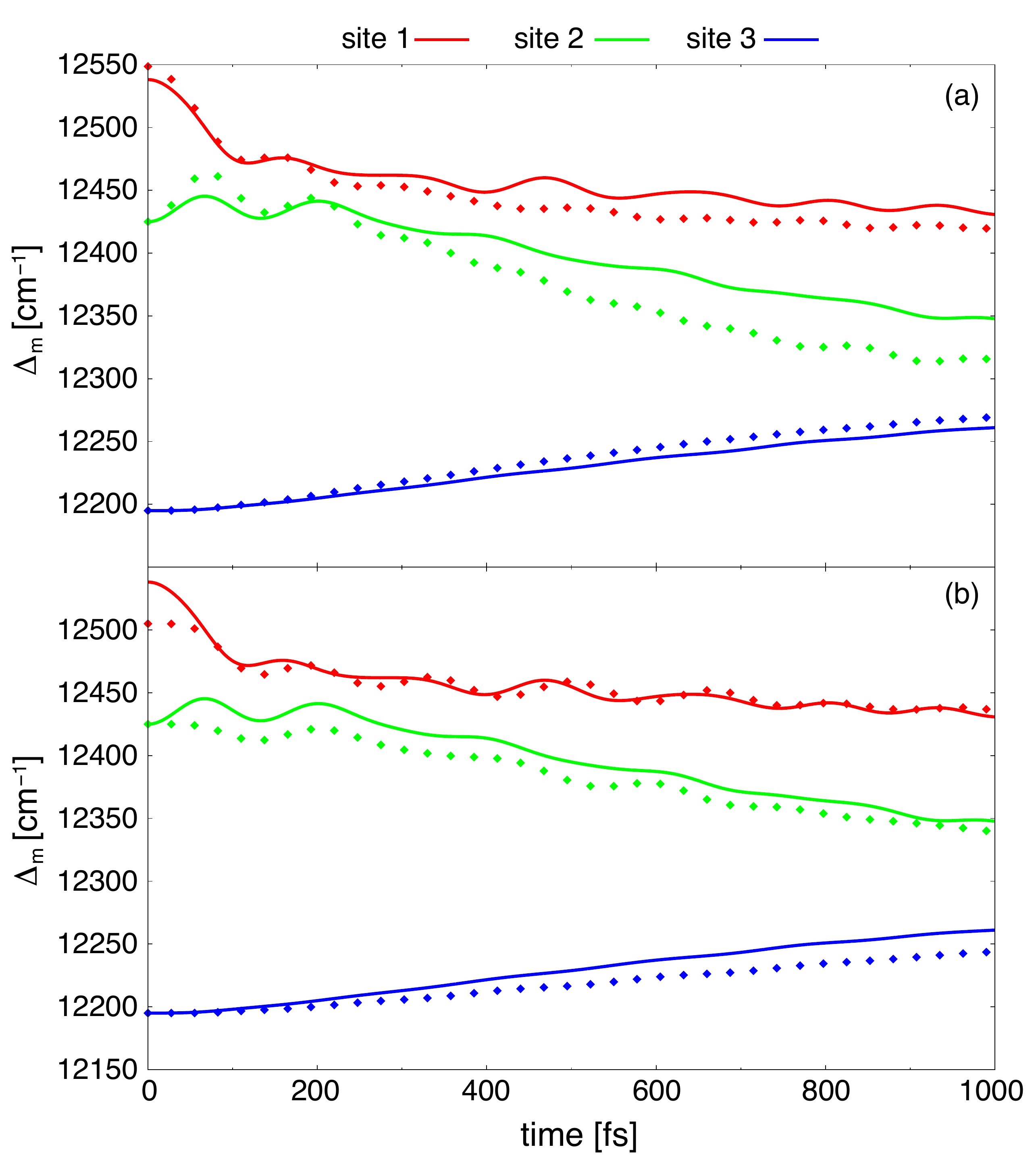}
\caption{
Energy gaps, $\Delta_m(t)$, according to Eq. (\ref{eq:gap}) for model I (solid lines) as compared with models II (a) and IV (b)  (dotted lines).}
\label{fig:gap}     
\end{figure}
%ffffffffffffffffffffffffffffffffffffffffffffffffffffffffffffffffffffffff

%ffffffffffffffffffffffffffffffffffffffffffffffffffffffffffffffffffffffff
\begin{figure}[t]
\includegraphics[width=0.475\textwidth]{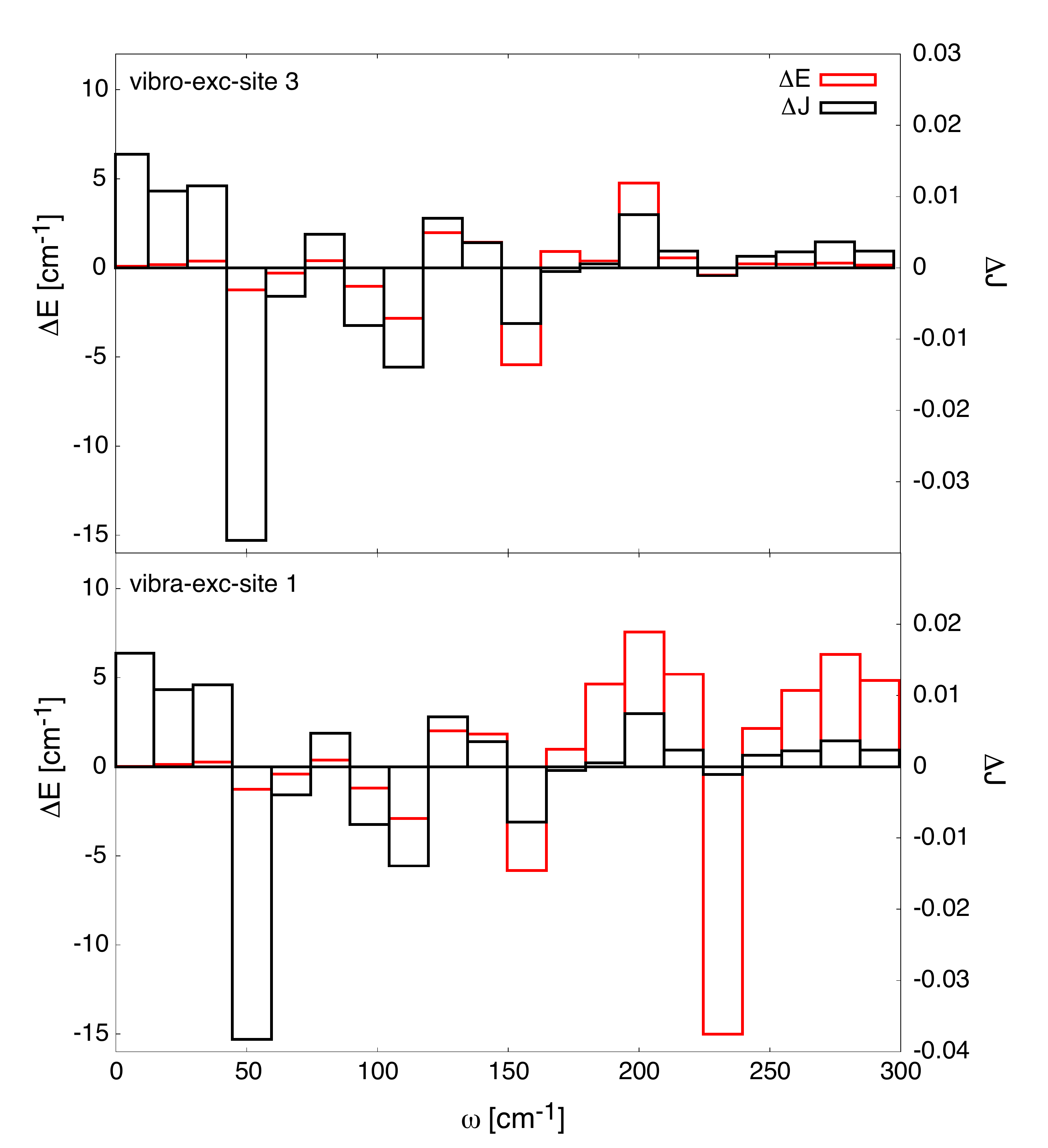}
\caption{
Difference ($\Delta E$) between vibronic energies for site 3 (upper panel) and vibrational energies at site 1 (lower panel) for simulations using the  Wendling (model I) and the Coker site 3 (model IV) SDs averaged with respect to the propagation time. The differences between the respective SDs are also shown ($\Delta J$). Data have been obtained by binning the frequency interval using a width of 15~\cm.
}
\label{fig:delta}     
\end{figure}
%ffffffffffffffffffffffffffffffffffffffffffffffffffffffffffffffffffffffff

\section{Results and Discussion}%%%%%%%%%%%%%%%%%%%%%%%%%%%%%%%%%%%%%%%%%%%%%%%%%%%%%%%%%%%
\label{sec:results}
 In Figs.~\ref{fig:wendling} to \ref{fig:coker_rescaled_site3} exciton population and vibrational as well as vibronic dynamics are presented for the four models. First, we discuss the population dynamics  taking model I as a reference. In all cases the diabatic populations show a beating between sites 1 and 2 with decreasing amplitude. Both state populations decay into state 3, whose population  increases almost monotonously. The four models differ in the amplitude of the oscillations, their decay as well as in the overall decay towards site 3. In particular, as compared to model I the oscillations are less pronounced in model II and more pronounced in models III and IV. The accumulation of population at the final site 3 is fastest in model II. For instance, the final population is about 0.67 and 0.8 in model I and II,  respectively. Models III and IV behave similar to model I in this respect.

 The adiabatic populations do not show an oscillatory behavior, instead they reflect a decay of the states at higher energy towards the lowest energy state. The latter is almost identical to the diabatic state 3 such that diabatic and adiabatic populations are rather close to each other. As a consequence the acceleration of the dynamics for model II is seen in both representations. The different models can be further distinguished by means of two characteristics of the populations dynamics. These are the behaviors of $P_{\alpha=2}$ and $P_{\alpha=3}$, which signal how long population is trapped in the highest and intermediate excited state. 
 
The exciton population relaxation towards the lowest state is due to energy dissipation into the vibrational DOF. The associated time scale should be governed by the strength of EVC, i.e. the HR factor. Thus, it is not surprising that model II shows the fastest population of the lowest state, whereas models I, III, and IV, which have the same total HR factor, behave similar. However, inspecting the dynamics of the higher excited states, we notice that there is a difference, which doesn't have an obvious relation to the total HR factor. For instance, the population of the highest adiabatic state, $P_{\alpha=3}$,   becomes close to zero in the order model I, II, III, IV. Further, at the end of the propagation interval the values of the intermediate state population $P_{\alpha=2}$ are in the order model III, I, IV, II. From this difference between I and III/IV we conclude that although the transfer rate through the complex is governed by the total HR factor, it is the shape of the SD which determines the details of the relaxation dynamics.
      
 Next, we focus on the vibrational and vibronic dynamics in the local potential energy surfaces. The behavior of model I has been rationalized previously in terms of two basic mechanisms~\cite{schulze16_185101}. The extent of vibrational excitation in the electronic ground state of sites 1 and 2 is due to the competition between wavepacket motion in the electronic excited state and exciton transfer. Upon transfer, the wavepacket will be projected back onto the electronic ground state. Here, its displacement away from the equilibrium position is the larger the shorter the vibrational period is with respect to the transfer time. Therefore, there appears to be an almost sharp cut-off at around 160~\cm below which vibrational periods are below the transfer time. In contrast, the vibronic excitation at site 3 can be traced to vibrationally-assisted transfer, i.e. the narrow range of excited modes just provides good resonance conditions to compensate for the mismatch between energies of sites 2 and 3. 
 
 This behavior is essentially recovered for all four models as can be seen in Figs.~\ref{fig:wendling} to \ref{fig:coker_rescaled_site3}. In fact, the differences between the models  relate just to the actual distribution and magnitude of vibrational and vibronic excitations. Here, it is the fact that the Coker SD is essentially dominated by a few intramolecular mode that gives rise to most differences. In this respect it is interesting to compare the averaged excitation energies, ${\mathcal E}_m(t)$, which are rather similar in shape for the different models. However, they appear to be shifted with respect to each other. 
 
 How this influences the transfer dynamics can be scrutinized by inspecting the local energy gaps, Eq.~(\ref{eq:gap}), shown in Fig.~\ref{fig:gap} for models I, II, and IV. Overall, the behavior of the different models is rather similar.  During the first 200~fs $\Delta_1$ and $\Delta_2$ approach each other, thus facilitating efficient energy transfer between these two sites and thus population switching. The decrease of $\Delta_1$ is essentially due to vibrational excitation at site 1, whereas the increase of $\Delta_2$ is due to vibronic excitation at site 2. Subsequently, $\Delta_1$ levels off and the difference between $\Delta_2$ and $\Delta_3$ gradually decreases, such that efficient transfer to site 3 becomes possible. This is due  to vibrational excitation at site 2 and vibronic excitation at site 3. Inspecting models I and II (panel a) one notices that the essential difference is in $\Delta_2$, which decreases more rapidly in model II as compared with model I. As far as models I and IV are concerned, Fig.~\ref{fig:gap}b shows that the difference $\Delta_2-\Delta_3$ is almost the same and so is the final population at site 3. However, $\Delta_1-\Delta_2$ is larger for model IV as compared with model I, which explains the more rapid depopulation of the initial state in the latter case.
 
Finally, we discuss in more detail how the differences in vibrational/vibronic excitation correlate with difference in the SD. In Fig.~\ref{fig:delta} differences in vibrational and vibronic excitation at site 1 and 3, respectively, are contrasted with differences in the SD for models I and IV. First, we notice that in the low-frequency region of the phonon wing up to about 100~\cm, differences in the SD don't really matter since the overall excitation level is very low. From 100 to about 150~\cm{} changes on SD come along with respective changes in vibrational/vibronic energy. In the range starting from about 150~\cm{}, however, noticeable effects can be observed. First, for the vibronic excitation only   narrow frequency ranges play a role (in accord with the above mentioned resonance-assisted transfer). Second, for vibrational excitation there are some frequency intervals where already small changes in SD have a large effect on the level of excitation. Judging this effect one should keep in mind that in the considered frequency range model IV has only discrete modes whereas in model I the SD is continuous. This way the HR factor for the interval 225-240~\cm{} of model IV (a single mode) is diluted over four modes in model I.
\section{Summary}%%%%%%%%%%%%%%%%%%%%%%%%%%%%%%%%%%%%%%%%%%%%%%%%%%%%%%%%%%%
\label{sec:summary}
The ML-MCTDH approach has been applied to study the dynamics of four different models describing the excitation energy transfer in the FMO complex. These models differed in their SDs, which were taken from experiment~\cite{wendling00_5825} and recent calculations~\cite{lee16_3171}. As compared with density matrix based approaches, ML-MCTDH has the advantage  that a high-dimensional wavepacket is propagated such that mode-specific information is available, while approaching the continuum limit for the vibrational DOFs. The simple form of the Frenkel exciton Hamiltonian with linear EVC, greatly facilitates the ML-MCTDH implementation and its numerical feasibility. In contrast to density matrix approaches, there appears to be no restriction as far as the actual form of the SD is concerned (see also, Refs.~\cite{schulze15_6211,schulze16_185101,seibt09_13475,shibl17_xxx}). 

For the specific problem of FMO energy transfer, the following main conclusions could be drawn: First, the total HR factor is more decisive for the rate of population trapping at the lowest energy site than the actual shape of the SD. Second, the shape of the SD determines the distribution of vibrational and vibronic excitations and thus the local energy gap. Even for identical total HR factors, energy gaps may differ and thus the actual pattern of population relaxation. For instance, depending on the SD one may observe transient trapping of intermediate state populations. 

As a consequence, care must be taken when comparing different SDs. Even though they might look similar at first glance, small differences might become amplified if they occur in a frequency range relevant for the quantum dynamics.
\section{Acknowledgments}%%%%%%%%%%%%%%%%%%%%%%%%%%%%%%%%%%%%%%%%%%%%%%%%%%%%%%%%%%%
This work was made possible by NPRP grant \#NPRP 7-227-1-034 from the Qatar National Research Fund (a member of Qatar Foundation). The statements made herein are solely the responsibility of the authors.
One of the authors (O.K.) gratefully acknowledges practical help with the MCTDH code and many stimulating discussions with H.-D. Meyer (Heidelberg) during the last 17 years.
%
%\bibliographystyle{elsarticle-num}
%\bibliography{B_FMO-MCTDH-3}

%
\end{document}